\documentclass{emulateapj}

\def\ms{m~s$^{-1}$}
\def\ks{km~s$^{-1}$}
\def\msini{$M_P\sin{i}$}

\def\vsini{$V_{\rm rot}\sin{i}$}
\def\teff{$T_{\rm eff}$}
\def\msun{M$_{\odot}$}
\def\mjup{$M_{\rm Jup}$}

\def\feh{[Fe/H]}

\def\nstar{1194}
\def\abest{1.0} 
\def\bbest{1.2}
\def\cbest{0.07}

\begin{document}
\title{Giant Planet Occurrence in the Stellar Mass-Metallicity Plane}

\author{John Asher Johnson\altaffilmark{1,2},
Kimberly M. Aller\altaffilmark{2,3},
Andrew W. Howard\altaffilmark{3,4},
Justin R. Crepp\altaffilmark{1} 
}

\email{johnjohn@astro.caltech.edu}

\altaffiltext{1}{California Institute of Technology,
  Department of Astrophysics, MC 249-17, Pasadena, CA 91125; NASA
  Exoplanet Science Institute}
\altaffiltext{2}{Institute for Astronomy, University
  of Hawaii, Honolulu, HI 96822} 
\altaffiltext{3}{Department of Astronomy, University of California,
Mail Code 3411, Berkeley, CA 94720} 
\altaffiltext{4}{Townes Fellow, Space Sciences Laboratory, University
  of California, Berkeley, CA 94720-7450 USA}

\begin{abstract}
Correlations between stellar properties and the occurrence rate of
exoplanets can be used to inform the target selection of future planet
search efforts and provide valuable clues about the planet formation
process. We analyze a sample of 1266 stars drawn from the California
Planet Survey targets to 
determine the empirical functional form describing the likelihood of a
star harboring a giant planet as a function of its mass and
metallicity. Our stellar sample ranges 
from M dwarfs with masses as low as 0.2~\msun\ to intermediate-mass
subgiants with masses as high as 1.9~\msun. In agreement with previous
studies, our sample exhibits a
planet-metallicity correlation at all stellar masses; the fraction of 
stars that harbor giant planets scales as $f \propto
10^{\bbest \rm [Fe/H]}$. We can rule out a
flat metallicity relationship among our evolved stars (at 98\%
confidence), which argues that the high metallicities of stars with
planets is not likely due to convective envelope ``pollution.'' Our
data also rule out a constant planet  
occurrence rate for [Fe/H]~$< 0$, indicating that giant planets
continue to become
rarer at sub-Solar metallicities. We also find that planet occurrence
increases with stellar mass ($f
\propto M_\star$), characterized by a rise from 3\% around M dwarfs
(0.5~\msun) to 14\% around A stars (2~\msun), at Solar metallicity. We
argue that the correlation between stellar properties and giant planet
occurrence is strong supporting evidence of the core
accretion model of planet formation. 
\end{abstract}


\section{Introduction}
Mass and chemical composition are key
quantities in the formation, evolution and fate of stars. A star of a
given age is, to first order, characterized by these two physical
parameters, and the influences of mass and metallicity extend to the
formation and evolution of planets \citep{johnson09rev}. Even the
first handful  
of exoplanet discoveries revealed that the likelihood of a star harboring
a planet was closely tied to stellar iron content, or metallicity
\feh\ \citep{gonzalez97}. Subsequent studies of larger samples
of stars using uniform 
spectroscopic modeling techniques found that giant planet occurrence
increases sharply for stellar metallicity in excess of the Solar value,
rising from 3\% for [Fe/H]~$\lesssim 0$ to 25\% for [Fe/H]~$> +0.4$
\citep[][hereafter FV05]{santos04,fischer05b}.  

In addition to informing models of planet formation
\citep{ida05a,mordasini09,johansen09}, the planet-metallicity correlation (PMC)
has provided a guide for the target selection of subsequent planet 
searches. The {\it Next 2000 Stars} (N2K) and Metallicity-Biased CORALIE 
surveys leveraged the higher metallicities of their samples to detect
large numbers of close-in planets, many of which transit their host
stars and thereby yield key insights into the interior structures of 
Jovian exoplanets
\citep{fischer05a,bouchy05,johnson06,moutou06}. Indeed, studies
of known transiting planets have revealed 
evidence of a correlation between 
planetary core mass and the metallicity of their host stars
\citep{sato05,torres08,guillot06,burrows07c}.  

While the first planet detections yielded a definitive correlation
between giant planet occurrence and stellar metallicity, until
recently very 
little was known about the effects of stellar mass \citep{laws03}. The
first Doppler-based planet surveys concentrated primarily on stars
with masses similar to the Sun, both because it was desirable to find
Solar System 
analogs and because Sun-like stars make excellent
planet-search targets. Compared to more massive stars, dwarfs with
masses within  $1.0\pm0.2$~\msun\ are relatively
numerous, have cool atmospheres and slow rotational velocities
(\vsini~$ \lesssim 5$~\ks). The latter two features result in a high
density of narrow absorption lines in the spectra of Sun-like
stars, which is ideal measuring 
for stellar Doppler shifts to high precision. 

Stars at the
lower end of the mass scale (the K and M stars) are even more
numerous than the Sun and they also display large number of narrow
absorption features in their spectra. However, most low-mass stars are
optically faint ($V_{mag} \gtrsim 10$) and are thus not included in
large numbers in most 
Doppler surveys. The faintness of late-K and M-type dwarfs can be overcome
by using larger 
telescopes \citep{butler04,bonfils05b}, and more recently by observing
at infrared wavelengths \citep{bean10}. Despite the small numbers of M
dwarfs thus far monitored 
by Doppler surveys, one result has become apparent: M dwarfs
harbor Jovian planets very infrequently. Only eight systems containing
one or more giant planets have been found among the $\approx 300$ M
dwarfs on various Doppler programs \citep{johnson10a, nader10}. 

It was originally thought that
the paucity of Jupiter-mass planets around M  dwarfs was due to a
metallicity bias among nearby, low-mass stars
\citep{bonfils05a}. However, a recent study by \citet{johnson09b} 
revealed that M dwarfs likely have the same metallicity distribution
as Sun-like stars, and stars with masses $M_\star <
0.5$~\msun\ are 2-4 times less likely than Sun-like stars to 
have a Jupiter \citep{johnson07b, johnson10a}. 

At the other end of the mass scale, the problems inherent to
massive, early-type stars can be overcome by observing targets at a
later stage 
of their evolution \citep{hatzes03, setiawan05, sato05, reffert06,
  johnson07,nied07, liu08, dollinger09}. Once stars exhaust their core 
hydrogen fuel 
sources they move off of the main sequence, become cooler, and shed
a large fraction of their primordial angular momentum \citep{gray85,
  donascimento00}. The effects of 
stellar evolution transform a 2~\msun\ star from an A-type
dwarf with \vsini~$\sim 100$~\ks\ and \teff~$ = 8200$~K, to a K-type
subgiant or giant with \vsini~$< 2$~\ks\ and \teff~$\approx 4800$~K
\citep{demedeiros97, girardi02,sandage03}.  
Surveys of ``retired'' massive stars have resulted in the discovery of
$\approx 30$ Jupiter-mass planets with well-characterized orbits
\citep[see e.g. Table 1 of][]{bowler10}. 

Using a sample of stars spanning a wide range of masses,
\citet{johnson07b} measured a positive correlation between stellar
mass and the fraction of stars with detectable planets\footnote{In that
  study, 
  ``detectable planets'' were defined as having \msini~$ > 0.8$~\mjup\
  and $a < 2.5$~AU.}. In a related study, \citet{bowler10} measured a planet
occurrence rate of $26^{+9}_{-8}$\% among a uniform sample of 31
massive subgiants. Furthermore, based in part on a study of planets
around K giants in nearby open clusters, \citet{lovis07} found that
the average planet 
mass increases as a function of stellar mass, indicating that gas
giant planets become either more massive on average, or more numerous
(or both) with increasing stellar mass \citep[see also][]{bowler10}. 

The observed correlation between
stellar mass and the occurrence 
of detectable planets, like the PMC before
it, has 
added an important new variable to models of planet formation. While
the Sun and Solar-mass stars serve as important 
benchmarks for understanding the formation of our own planetary
system, successful, generalized planet formation theories must now
account for the effects of stellar mass, and presumably by extension,
disk mass \citep{laughlin04, ida05a, kennedy08, currie09}. 

While previous studies have uncovered the existence of a positive
correlation between stellar  
mass and planet occurrence, it is important to understand the
underlying functional form of the relationship. For example, it would
be advantageous to know whether the correlation is
purely linear, or if it can instead be better described as some
other functional form. Besides informing theories of planet formation,
an improved understanding 
of the relationship between planet occurrence and stellar mass will
also help guide the target selection of future surveys, and aid in the
interpretation of results of current and future planet-search efforts. Just
as some previous Doppler surveys biased their target selection toward
high-metallicity stars to increase their yield, future direct-imaging,
astrometric and Doppler 
surveys may benefit from concentrating on more massive stars.  This
strategy has paid off for one high-contrast imaging survey, resulting
in the detection of three giant planets around the A5 dwarf HR\,8799
\citep{marois08}. Another example of an imaged planet is Fomalhaut\,b,
which is a giant planet ($\lesssim3.3$~\mjup) orbiting just inside of
a debris disk of an A3V star \citep[][]{kalas08,chiang09}. Even in the
cases when surveys do not yield detections, proper interpretation of
null results requires knowledge of the expected
number of detections \citep[e.g.][]{nielsen09}.  

Ascertaining the underlying form of the dependence of giant planet
occurrence on stellar mass requires a larger sample than used in previous
studies. Since the publication of \citet{johnson07b} a
sample of 240 new intermediate-mass subgiants have been added to the
California Planet Survey (CPS) at Keck Observatory
\citet{johnson10b}. At the low-mass end, two new giant planets have
been  discovered among the CPS Keck sample of M dwarfs
\citep{johnson10a}. Improvements in 
our ability to estimate the metallicities of M dwarfs and
massive evolved stars have provided vital information about how to
properly isolate the effects of stellar mass from the known effects of
stellar metallicity. With these tools at hand, we are now poised to
make an updated evaluation of the relationship between stellar mass
and planet occurrence. 

Our paper is organized as follows. In \S~\ref{sample} we present the
characteristics of our three primary samples, including low-mass M
dwarfs from Keck observatory; ``Sun-like'' late-F, G and K (FGK) dwarfs
from the main CPS sample; and massive, evolved stars from the Lick and
Keck subgiant 
surveys. In \S~\ref{metallicity} we examine the separate effects of
mass and metallicity on planet occurrence. In \S~\ref{fraction}
we present our Bayesian inference technique of
measuring correlations between planet occurrence and stellar
characteristics and we provide the
best-fitting parameters for the measured relationship in
\S~\ref{results}. We compare our results 
with previous work in \S~\ref{comparison}. Finally, we summarize our
key results and discuss our findings in the 
context of the current theoretical understanding of planet formation
in \S~\ref{summary}. 

\begin{figure}
  \epsscale{1.2}
    \plotone{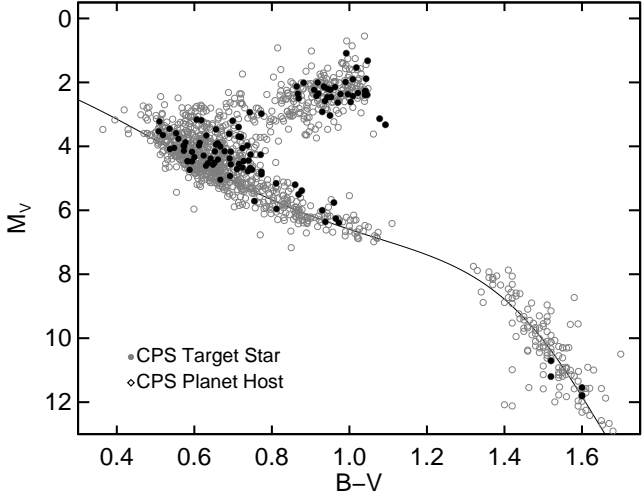}
     \caption{H-R diagram showing our stellar sample (filled circles)
       and planet host stars (open circles). The solid line is the
       polynomial relationship describing the mean \emph{Hipparcos}
       main sequence \citet{wright04}. The excess scatter seen about the
       lower main sequence $B-V > 1.2$ is primarily due to errors in
       the published  V-band magnitudes and/or parallaxes for those
       faint stars. \label{fig:hr}} 
\end{figure}

\section{Selection of Stars and Planets}
\label{sample}

Our goal is to measure planet occurrence as a function of stellar
properties. Care must be exercised in selecting
the sample of target stars such that planets of a given
mass and orbital semimajor axis could be uniformly detected over the
entire sample. The criteria for planet mass 
and semimajor axis translate into limits on velocity amplitudes, $K$,
and orbital periods that can be tallied among the
sample of planet detections. These criteria must be selected to ensure
reasonably uniform detection characteristics across several Doppler
surveys, which have different detection sensitivities and time
baselines.  

In what follows, we describe our selection of stars and planets from
among the various CPS planet search programs. The CPS is a
collection of Doppler surveys carried out primarily at the Lick and
Keck Observatories. The CPS target lists
provide a large stellar sample with a wide
range of masses and metallicities. Specifically, our stars lie in the
ranges  $0.2 < M_\star/M_\odot \lesssim 2.0$ and $-1.0 < $~\feh~$<
+0.55$.  The long 
time baselines ranging from 3 to 10 years, and Doppler precision
ranging from 1--5~\ms\ have 
resulted in a diverse and fairly complete sample of giant planets
that have been compiled in the Catalog of Nearby Exoplanets (CNE,
\citet{butler06}) as updated by Wright et al. 2010 (in
prep) in the Exoplanet Orbit Database\footnote{{\tt
    http://exoplanets.org/}}.  

\begin{figure*}[!t]
  \epsscale{1.}
    \plotone{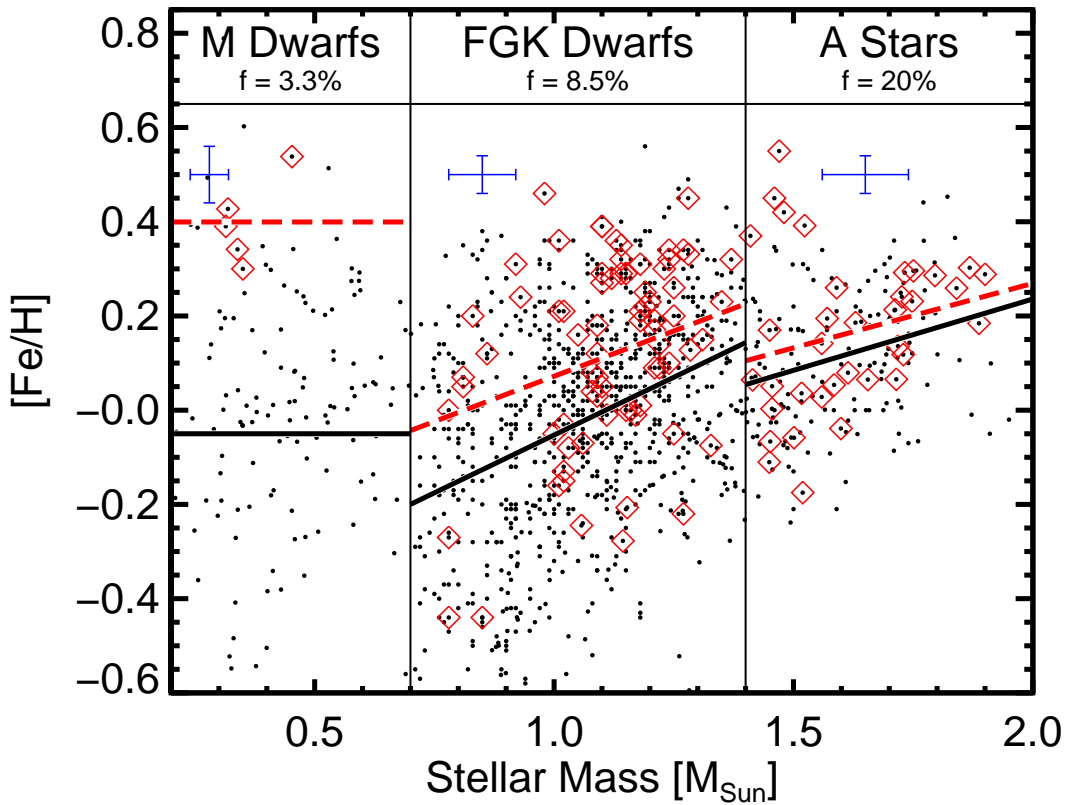}
     \caption{A plot of stellar mass ($M_{\star}$) and metallicity
      ([Fe/H]) for the full stellar sample, comprised of \nstar\
      stars (black dots), 115 of which harbor at least one detectable
      planet (red diamonds). For
      visualization, we have divided the stellar sample into
      three       broad groups: M dwarfs, FGK dwarfs, and massive
      ``retired'' A stars. The fraction $f$ of stars with planets is
      printed above each group. The thick (black) lines each stellar mass group
      represent the best-fitting linear relationships between mass and
      metallicity (for the M dwarfs the lines represent the
      metallicity). The dashed (red) line is the best-fitting linear 
      relationship 
      between mass and metallicity for the stars with planets. For the
      M dwarfs we simply report the average metallicity for each
      population. The (blue) 2-dimensional error bars represent the
      typical measurement uncertainties. In each 
      mass group, there is a systematic metallicity offset between the
      stars with and without planets. Discontinuities between the
      samples are not entirely physical, and are in large
      part due to the different target-selection criteria for the 
      three surveys.  \label{fig:massmet}}
\end{figure*}

\subsection{Stellar Sample}

The low-mass stars in our sample are drawn from from the CPS Keck
survey of late-K and M-type dwarfs 
\citep{rauscher06, johnson10a}. This sample comprises 
stars with $M_\star < 0.6$~\msun\ as estimated with the photometric
calibration of \citet{delfosse00}. We estimate the metallicities with
the broadband 
photometric calibration of  \citet{johnson09b}, which
relates the metallicity of a star to its ``height'' ($\Delta M_K$)
above the mean main-sequence in the $\{V-K_S$, $M_{K_S}\}$
plane. 

The bulk of our Solar-mass F, G and K dwarfs are taken from
the Spectroscopic Properties of Cool Stars catalog
\citep[SPOCS;][]{valenti05}. Most of these stars have masses
in the range $0.8 < M_\star/M_\sun < 1.2$. However, the SPOCS catalog
contains some higher mass subgiants, which we fold into our high-mass
stellar sample described herein.   
The spectroscopic properties listed in the SPOCS catalog were measured
using the LTE spectral synthesis software package {\it Spectroscopy
  Made Easy} \citep[SME;][]{valenti96}, as described by 
\citet{valenti05} and FV05. Stellar masses for the SPOCS
catalog are cataloged by \citet{takeda07}, who 
associate the spectroscopic stellar properties to isochrones computed
using the Yale Stellar Evolution Code \citep[YREC][]{yrec}. 

We select our high-mass stellar sample from the Lick and Keck Subgiant
Planet Surveys. The sample selection is described in
\citet{johnson06} and \citet{johnson10b}. The masses
and metallicities of the subgiants in our sample are estimated using
SME and are listed
in the fourth contribution to the SPOCS catalog (Johnson et al. 2010c, 
submitted). The majority of 
our subgiants have masses in the range 1.3--2.0~\msun, with a tail in
the distribution extending to 1.0~\msun. The metallicities of the
subgiants range from \feh~$=-0.2$ to $+0.5$. 

Our full stellar sample contains \nstar\ stars: 142 M and late-K
dwarfs from the Keck 
M Dwarf Survey \citep{butler06b}, 807 dwarf and subgiant stars from
the original SPOCS catalog, and 246 subgiants from the SPOCS
IV. catalog. Figure~\ref{fig:hr} shows our stars in the $\{M_V, B-V\}$
  H--R diagram. The open symbols are the positions of all of our
  stars, and the filled symbols are the stars known to harbor at least
  one detectable (giant) planet, as described in the following section.
\\

\subsection{Planet Detections}

Following FV05, we restrict our analysis to systems with
at least one ``uniformly detectable planet,'' which we 
define as those with velocity semiamplitudes $K > 20$~\ms\ and
semimajor axes $a < 2.5$~AU. We
decreased threshold in $K$ from the value used by 
FV05 ($K > 30$~\ms) because of the increased Doppler
precision of HIRES 
since the 2005 detector upgrade \citep[see e.g.][]{howard10a}. For
reference, at 1~AU and for circular orbits, semiamplitudes $K =
20$~\ms\ corresponds to minimum planet masses 
\msini/~\mjup~$ = \{0.44, 0.82, 1.12\}$ for
M$_\star/$\msun~$=\{0.4,1.0,1.6\}$. 

Due to the limited time  
baselines of the Doppler surveys from which our targets are
drawn, we also restrict our analysis to planets with $a <
2.5$~AU. This criterion is set primarily by our sample of
intermediate-mass subgiants, which are on surveys with time baselines
ranging from 3--6 years.  These criteria will, for most stellar
masses, represent conservative cuts on the total number of {\it
  giant} planet detections.  We defer the
analysis of the frequency of less massive  planets with
\msini~$<1.0$~\mjup\ or orbits wider than 2.5~AU to other studies
\citep[e.g.][]{sousa08, howard09, gould10, cumming08}. 

To further ensure uniform detectability within our stellar sample we
restrict 
our analysis to stars with a minimum number of observations. For the
low-mass and Solar-mass samples we require $> 10$ observations. For
the high-mass subgiants we require $> 6$ observations; a smaller number
owing primarily 
to the shorter time baseline of our Keck survey. We also require 
minimum observational time baselines corresponding to our semimajor 
axis limit of $a < 2.5$~AU. Thus, for the M dwarfs we require a
baseline $> 6.3$~years, using an average stellar mass $M_\star =
0.4$~\msun; $> 4$ years 
for the Solar-mass stars; and $> 3$ years for the subgiants 
with an average stellar mass of $1.6$~\msun.

We compiled our sample of planet detections by cross-correlating our
stellar samples with the Exoplanets Data Explorer, and recent planet
announcements from the 
CPS  \citep{howard10b,johnson10b,johnson10a}. We augmented this list
of secure 
detections with unpublished detections from the Keck Subgiants Planet
Survey. These unpublished candidates all have more
than 10 observations over 
$\approx 3$ years, but lack strong enough constraints on the orbital
parameters for publication. However, since the present study is
concerned with planet 
\emph{occurrence} we feel confident in including these secure, yet
unpublished detections in our sample. All of the unpublished
candidates have radial velocity variations consistent with Doppler
amplitudes and periods that meet our criteria for uniform
detectability.

Our sample of planet detections comprises 5 planets around M dwarfs,
74 planets around the SPOCS sample of FGK dwarfs, and 36
planets around subgiants. \\

\section{Disentangling Mass and Metallicity}
\label{metallicity}

In our analysis 
we treat stellar mass and metallicity as separate independent
variables affecting the likelihood that a star harbors a planet. The
validity 
of this premise rests in part on the analysis of FV05,
who 
noted an artificial correlation between mass and metallicity in the
SPOCS sample that is due to the color and magnitude cuts used in the
target selection: the more massive stars in the 
SPOCS sample have higher metallicities than the lower-mass stars
\citep[][;FV05]{santos04,marcy05b}. This 
selection effect is clearly seen in our updated data set shown in the
middle panel of Figure~\ref{fig:massmet}. However, as can 
be seen in that figure and as noted by FV05,
there is a metallicity offset between stars with and without planets
at all masses between 0.7~\msun\ and 
1.4~\msun (see also \citet{santos04}). Thus, despite the artificial
mass-metallicity correlation 
in our sample of FGK dwarfs, there still exists a clear 
PMC. At a given mass, stars with planets have higher
metallicities than the stars without planets. 

The PMC is also apparent in
the M dwarf sample. The low-mass stars with planets are
extremely metal-rich compared to the full stellar
sample\footnote{The metallicities of the full sample of M dwarfs were
  estimated using the photometric calibration of Johnson \& Apps
  (2009). This required an extrapolation of their relationship for
  the stars below the main sequence. However, since the relationship
  between $\Delta M_K$ and [Fe/H] is expected to be monotonic, our
  extrapolation will not affect our conclusions in this case.}.  
Also apparent from the M dwarf sample is that there are far fewer
planet detections, both in an absolute and fractional sense, compared
to the higher-mass stellar samples. 

The far right-hand panel of  Figure~\ref{fig:massmet} shows that the
metallicity offset between stars with and without planets is also
present among the more massive subgiants, albeit at lower
statistical significance. Like the FGK dwarfs, the subgiants have a
artificial mass-metallicity correlation, 
owing to the red cutoff of ($B-V < 1.1$) used in the selection of the
subgiants from the \emph{Hipparcos} catalog. A much higher fraction of
massive stars have detected planets than do the M or FGK dwarfs. 

The metallicity offsets among the stars with and without
planets in the three mass-bins in
Figure~\ref{fig:massmet} are suggestive of a PMC that spans an order of
magnitude in mass, from 0.2~\msun\ to 2.0~\msun. Also seen among the
three mass bins is a steadily increasing planet occurrence rate: while
only 3.3\% of the M dwarfs have a planet, 20\% of the retired A
stars harbor one or more giant planets. This is strong evidence that
planet occurrence correlates with stellar mass, separately from the
effects of stellar metallicity. In the following sections we examine
these trends in further detail.

\section{Quantifying Planet Occurrence}
\label{fraction}

\subsection{Parametric Description}

We derive a parametric relationship between stellar
properties and fraction of stars with planets using Bayesian
inference. The resulting function, while ad hoc, can be 
used to predict yields of future planet surveys, interpret the
results of ongoing planet search efforts, and compared
directly to the output of theoretical models of planet formation. 

Our choice of functional form follows from the metallicity analyses of
FV05 and \citet{udrysantos07}, who describe the fraction 
of stars with 
planets, $f$, as a function of metallicity in the form $f(F) \propto
10^{\beta F}$, where $F \equiv {\rm [Fe/H]}$. Our parametric model
also needs to account for stellar mass. Previous 
observational studies suggest that planet occurrence should rise
monotonically with stellar mass \citep{laws03, johnson07b,
  lovis07}. For the mass relationship we adopt a 
power law $f(M) \propto M^\alpha$, where $M \equiv
M_\star/M_\odot$. 

Since we assume that mass and metallicity produce
separate effects, the fraction of stars with 
planets as a function of mass and metallicity can be described by

\begin{equation}
\label{eqn:model}
f(M,F) = C M^\alpha 10^{\beta F} 
\end{equation}

We note that there exist many possible functional forms for $f(M)$ and
$f(F)$. Indeed, any monotonic function should provide an adequate fit
to our data set. For example, \citet{robinson06} use a logistic function to
describe planet fraction as a function of stellar $\alpha$-element
abundance, and they note that a power law is simply an approximation
to the low-yield tail of such a function. However, we have decided to
use power law descriptions\footnote{Since [Fe/H]~$\propto
  log{N_{Fe}}$, the exponential term in Equation~\ref{eqn:model} is a power
  law relationship of the number of iron atoms: $f(F) \propto
  10^{\beta F} \propto N_{Fe}^\beta$.} due to the simplicity of the
functional form and for ease of comparison with previous studies.

\subsection{Fitting Procedure}
\label{fitting}

For conciseness, we denote the parameters in
Equation~\ref{eqn:model} by $X$. The parameters can be inferred from
the measured number of planet hosts $H$ drawn from a larger
sample of $T$ targets using Bayes' theorem: 

\begin{equation}
P(X \,| \,d)  \propto 
    P(d \,| \,X) P(X) 
\label{eqn:bayes}
\end{equation}

\noindent where $P(X \,| \,d)$ denotes the
probability of $X$ conditioned on the data $d$. In our analysis, the
data represent a binary result: a star does or does not have a
detectable planet. The 
terms on the right of the proportionality are the probability of the
data conditioned 
on the distribution of possible $X$, multiplied by the
prior knowledge and assumptions we have for the parameters.

Each of the $T$ target stars represents a Bernoulli
trial, so the probability of finding a planet at a given mass and
metallicity is given the binomial distribution. The probability of a
detection 
around star $i$ (of $H$ total detections) is given by $f(M_i, F_i)$. The
probability of the $j$th nondetection is $1 - f(M_j, F_j)$. Thus,

\begin{eqnarray}
P(X \,| \, d)  & \propto & P(X) \prod_{i}^{H}
f(M_i,F_i) \nonumber \\
& \times & \prod_{j}^{T-H}[1 - f(M_j, F_j)]
\label{eqn:product}
\end{eqnarray}

For each detection or nondetection, our measurements of the stellar
properties of each system, $M_i$ and $F_i$, are themselves probability
distributions given by $p_{\rm obs}(M_i, F_i)$. We approximate these
pdfs as the product of Gaussians\footnote{Because stellar
  metallicities are used to 
  select 
  the appropriate stellar model grids (``isochrones'') for the estimate
  of the stellar mass, these two measurements are actually
  covariant. However, we find that our result is not affected by
  assuming independent Gaussians.} with means $\{M_i, F_i\}$
and standard deviations 
$\{\sigma_{M,i}, \sigma_{F,i}\}$. The predicted planet fraction for
the $i$th star can then be expressed as 

\begin{equation}
f(M_i,F_i) = \int\int p_{\rm obs}(M_i, F_i) f(M, F) dM dF
\end{equation}

For ease of calculation the products in
Equation~\ref{eqn:product} can be rewritten as 
the sum of log-probabilities, or the marginal log-likelihood
\\

\begin{eqnarray}
\mathcal{L} \equiv \log{P(d \,| \,X)} & \propto & 
\sum_{i}^{H}\log{f(M_i,F_i)} + \nonumber \\
& + & \sum_{j}^{T-H}\log{[1 - f(M_j,F_j)]} \nonumber \\
& + & \log{P(X)}
\end{eqnarray}

\noindent The parameters $X = \{C,\alpha,\beta \}$ are then
optimized by maximizing $\mathcal{L}$ conditioned on the data.

We perform our maximum-likelihood analysis by numerically evaluating
$\mathcal{L}$ 
on a 3-dimensional grid over intervals bounded by uniform priors on 
the parameters $\{C,\alpha,\beta \}$. In our case, the
priors simply define the integration limits on the marginal
probability density functions (pdf) of the parameters, e.g.

\begin{equation}
P(\alpha \,| \, d) = \int_{\beta_{\rm min}}^{\beta_{\rm max}} 
\int_{C_{\rm min}}^{C_{\rm max}}P(X
\,| \, d) d\beta dC
\label{eqn:marg}
\end{equation}

\noindent The ranges of
the uniform priors used in the analysis are 
listed in the second column of Table~\ref{tab:pars}. 

\begin{deluxetable}{lccc}[!h]
\tablecaption{Model Parameters \label{tab:pars}}
\tablewidth{0pt}
\tablehead{
  \colhead{Parameter}    &
  \colhead{Uniform} & 
  \colhead{Median} & 
  \colhead{68.2\% Confidence}   \\
  \colhead{Name}    &
  \colhead{Prior\tablenotemark{a}} & 
  \colhead{Value} & 
  \colhead{Interval}   
}
\startdata
$\alpha$ & (0.0, 3.0)    & \abest   & (0.70, 1.30)   \\ 
$\beta$  & (0.0, 3.0)    & \bbest   & (1.0, 1.4) \\ 
$C$      & (0.01, 0.15)  & \cbest  & (0.060, 0.08)    \\
\enddata
\tablenotetext{a}{We used uniform priors on our parameters between the
  two limits listed in this column.}
\end{deluxetable}

It might at first seem more appropriate to use a prior for $\beta$ from
the analysis of FV05, e.g. a Gaussian centered on $\beta = 2$, rather
than a uniform function. However, we decided against this choice of
prior because 
no confidence interval for $\beta$ was reported by FV05, and we could
not be certain that their value was truly representative of our data
due to fundamental differences in our methodology, as we discuss
in \S~\ref{results}. Similarly, no functional form for $f(M)$ was
reported by \citet{johnson07b}. 

However, our choice of a uniform prior
is not entirely uninformed. Based on previous studies, we felt
it was  safe to consider only monotonically increasing functions
($\alpha > 0$, $\beta > 0$), and for $\beta$ we chose a range that
encompasses the value measured by FV05. 

\section{Results}
\label{results}

The best-fitting parameters and their 68.2\% (``1-$\sigma$'')
confidence intervals are listed 
in the third and fourth columns of Table~\ref{tab:pars}. We estimated
the confidence intervals by measuring the 15.9 and 
84.1 percentile levels in the cumulative distributions (CDF)
calculated from the marginal pdf of each parameter
(e.g. Equation~\ref{eqn:marg}).  

\begin{figure}[!t]
\epsscale{1.2}
\plotone{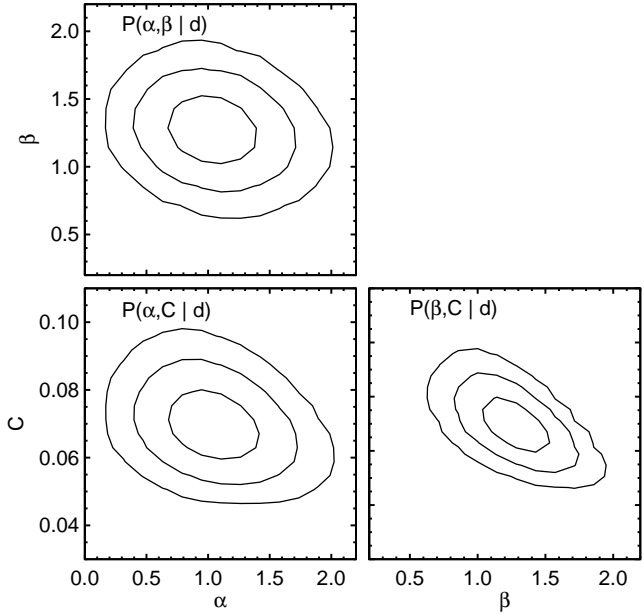}
\caption{Marginal posterior pdfs for the model parameters conditioned
  on the data. \label{fig:margcov}}       
\end{figure}

\begin{figure*}[!ht]
\plotone{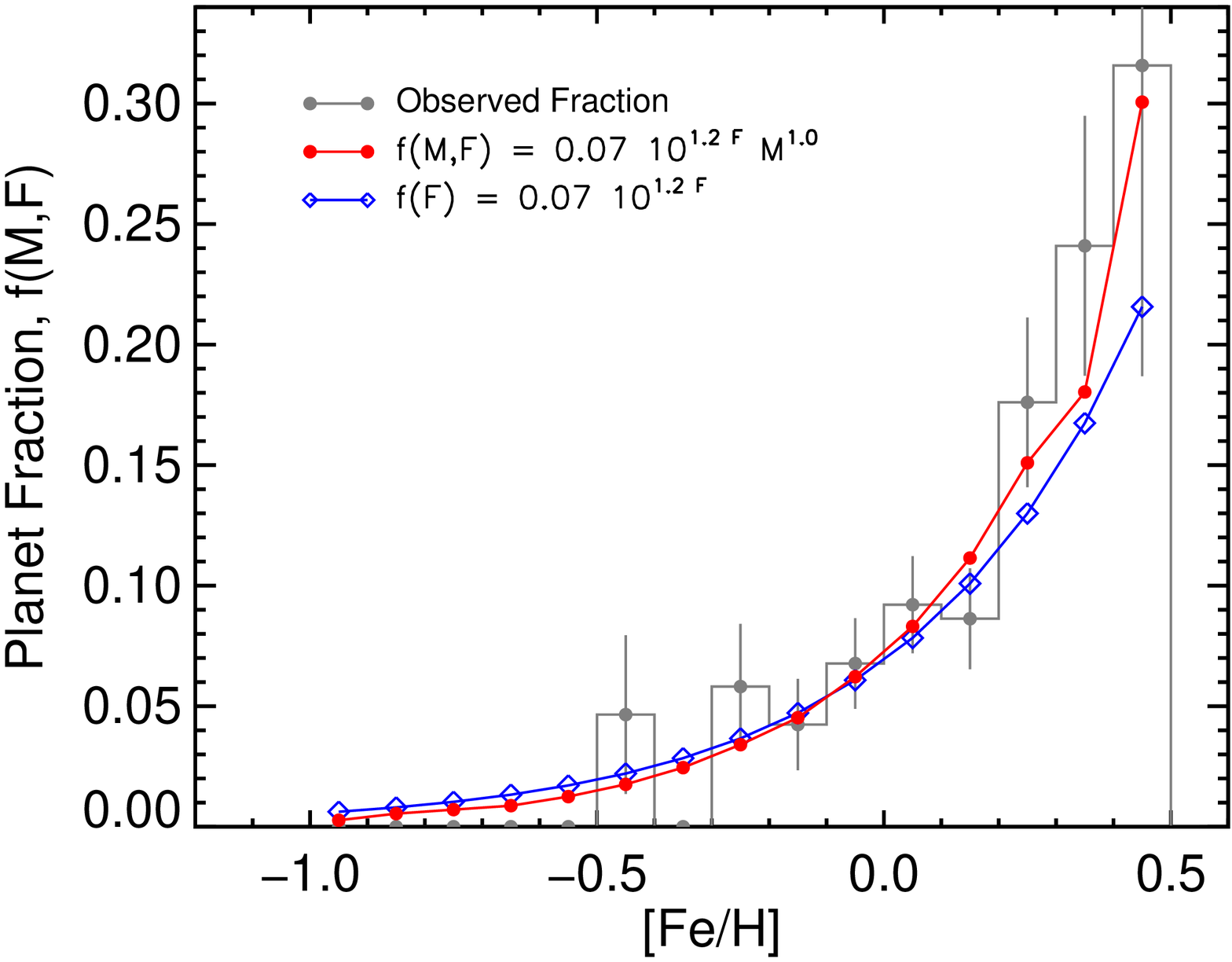}
\caption{Planet fraction ($f = N_{planets}/N_{stars}$) as a function
  of metallicity for our stellar sample (gray histogram). The red filled
  circles show the planet fraction predicted by
  Equation~\ref{eqn:model} for the masses and metallicities of the
  stars in each bin. Note that we use a histogram only for
  visualization purposes; the data were fitted directly without
  binning. The blue open diamonds show the best-fitting 
  relationship between planet fraction and stellar metallicity for
  $M_\star = 1$~\msun. None of the 52 stars with [Fe/H]~$<-0.5$ harbor
  a giant planet.
 \label{fig:modelfitF}}           
\end{figure*}

The marginal joint parameter pdfs are shown in
Figure~\ref{fig:margcov}. The comparisons between the best-fitting
relationship (Equation~\ref{eqn:model}) and the data are shown in
Figure~\ref{fig:modelfit} and \ref{fig:modelfitF}. In both figures,
the histograms show the ``bulk'' planet frequency, with bin widths of
0.15~\msun\ and 0.1~dex, respectively. The filled circles denote the
median planet fraction  
predicted by Equation~\ref{eqn:model} based on the masses and
metallicities of the stars in each bin. The diamonds show the
best-fitting metallicity and mass relationships, given by $f(M, F=0)$
and $f(M=1, F)$.

Our Bayesian inference analysis provides two additional assurances
that stellar mass and metallicity correlate separately with 
planet fraction. The first is the lack of covariance
between $\alpha$ and $\beta$ in Figure~\ref{fig:margcov}. This also
demonstrates that our stellar sample adequately spans 
the mass-metallicity plane despite the artificial correlation between
stellar parameters in part of our sample. The second check on our
initial assumptions 
is seen in Figure~\ref{fig:modelfit}. While some of the increase in
planet fraction as a function of stellar mass is due to a rise in
average stellar metallicity in our sample of high-mass stars
(circles), there still exists a nearly linear increase owing to
stellar mass alone (diamonds). Thus, there is an approximately
order-of-magnitude increase in planet occurrence over the mass range
spanning M dwarfs to A-type stars. Similarly, some of the metallicity
relationship is due to the higher stellar masses among the metal-rich
stars. However, there still exists a strong metallicity correlation
spanning more than an order of magnitude in iron abundance. 

In the following section we compare our results to those of related
studies. 

\section{Comparisons with Previous Work}
\label{comparison}

\subsection{Previous metallicity studies}

FV05 studied planet occurrence as a function of
metallicity among the Sun-like
portion of our stellar sample and found $\beta = 2$. By restricting
our analysis to the SPOCS subset of our sample that overlaps with
FV05, and by fitting a function of metallicity alone we find $\beta =
1.7 \pm 0.3$, which agrees with the FV05 value to within our 68.2\%
confidence interval. The significance of the difference is reduced
further if we assume the uncertainty in their measurement is
comparable to ours. 

By fitting for both mass and metallicity we find $\beta = 1.4 \pm 0.3$
and $\alpha = 0.7 \pm 0.4$, which agree with the values in
Table~1 measured for the full stellar sample. This provides 
assurance that our analysis is not overly sensitive to our high-mass
stellar sample, among which our detection sensitivity is 
lower due to the shorter time baseline and  fewer Doppler
measurements per star.  

It is likely that this smaller $\beta$ from our analysis
of the Sun-like stars compared to that of FV05 is in part due to the
different methods of 
fitting the  planet-fraction relationship, i.e. their least-squares
fit to histogram bins versus our Bayesian approach\footnote{For
  example, FV05 performed a $\chi^2$ minimization, which assumes
  symmetric ($\sqrt(N)$) error bars on their histogram 
  bins. However, the errors should have been binomial and
  asymmetric, which would have admitted smaller values of
  $\beta$.}. The other key difference is that we simultaneously
fit to both mass and metallicity. Since mass and metallicity are
correlated in our samples some of the metallicity relationship observed
by FV05 was due to stellar mass. This effect can also be seen in
Figure~\ref{fig:modelfit}. The joint mass-metallicity relationship
sits above the metallicity power-law at high values of [Fe/H] since
the metal-rich stars in our sample tend to be slightly more massive on
average than the metal-poor stars. 

\citet{udrysantos07} analyzed the FV05
sample, together with a sample of stars drawn from the CORALIE survey,
and  found $\beta = 2.04$ for [Fe/H]~$ > 0$. For lower metallicities
they suggest a flat occurrence rate provides a better fit than the
continuation of the exponential relationship to sub-Solar
metallicities. We compared the two functional forms (exponential
versus exponential-plus-constant) using the method of Bayesian model
comparison. By integrating the right-hand side of
Equation~\ref{eqn:bayes} over all parameters $X$, one obtains the
evidence, or total probability of the model conditioned on the data:

\begin{equation}
P(d) = \int \int \int P(\alpha, \beta, C\,|\,d)\, d\alpha\, d\beta\, dC
\end{equation}

\noindent The ratio of evidences provides a means of quantifying
preference in one model over another. If a model has evidence more
than a factor of 10 greater than the alternative, it is ``strongly
preferred'' \citep{bayesfactors}. 
When fitting planet--fraction as a function of
metallicity alone, the evidence for the exponential-only model is a
factor of 1800 higher than the exponential-plus-constant. {\it Thus,
  our data strongly prefer a model in which the fraction of stars with
  planets continues to decrease for [Fe/H]~$ < 0$.} 

We can take the Bayesian evidence analysis a step further and compare
our joint fit to the planet fraction as a function of mass and
metallicity to previous fits to metallicity or mass alone. We find
that the evidence for the joint fit is a factor of 2400 larger than
that of the metallicity-only fit, and a factor of $10^7$ higher than a
mass-only fit. The planet fraction among our sample is therefore best 
described as a function of metallicity {\it and} stellar mass. 

\subsection{Previous mass studies}

\citet{johnson07b} used
roughly the same sample presented herein to measure the occurrence
rate of planets in three coarse mass bins with 
widths of 0.6~\msun\ centered on $M_\star = \{0.4, 1.0,
1.6\}$~\msun. In these three intervals they measured occurrence rates
of $1.8\pm1.0$\%, $4.2 \pm 0.7$\% and $8.9 \pm 2.9$\%. After
correcting for the average stellar metallicity in each bin, the 
fractions change slightly to $2.5 \pm 1.2$\%, $3.5 \pm 0.7$, and a
lower limit of 6.3\% for the high-mass bin. Integrating our
relationship over the same mass intervals yields $2.5 \pm 0.9$\%, $6.5
\pm 0.7$\%, and $11 \pm 2$\%. 

\begin{figure*}[!ht]
\plotone{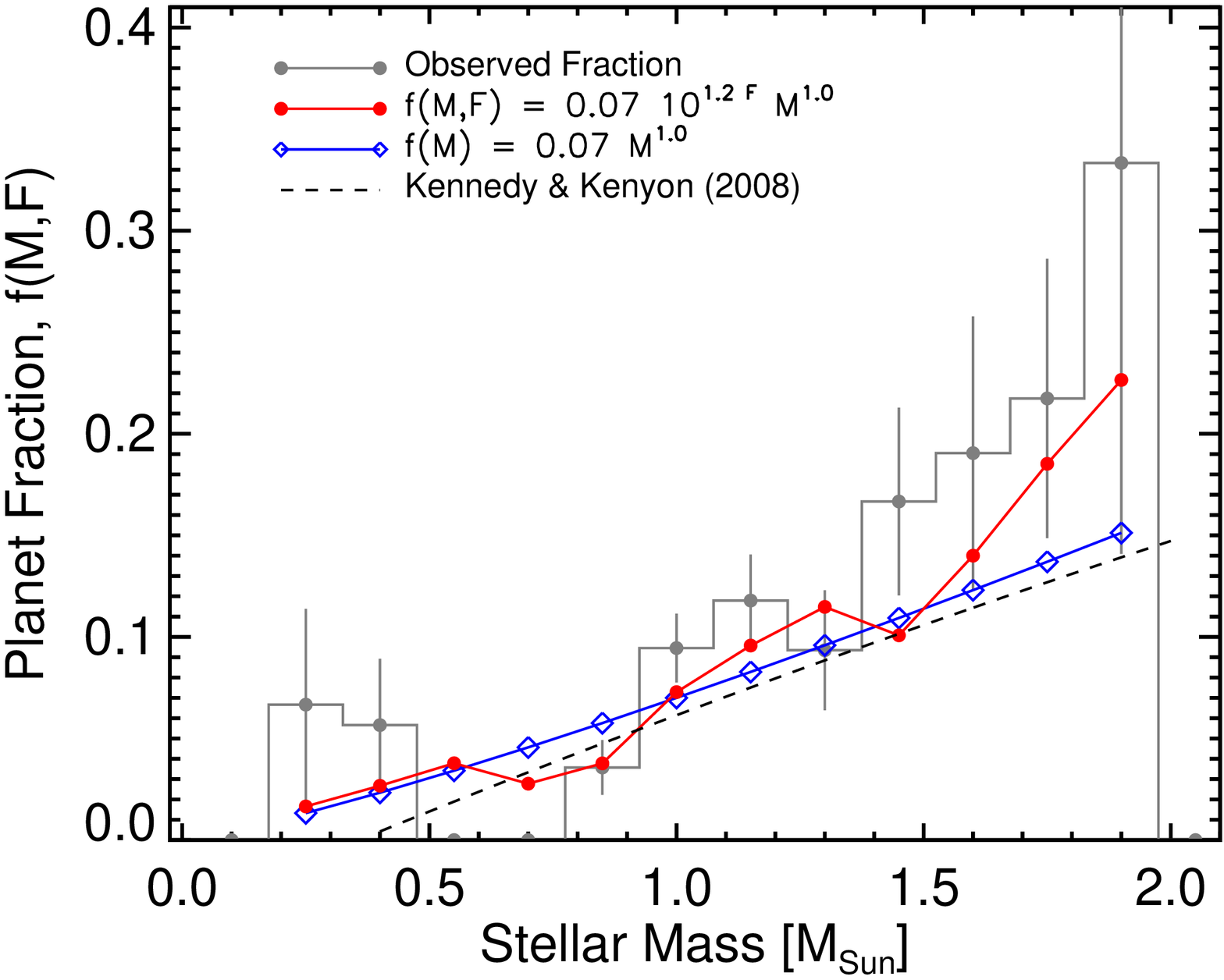}
\caption{Planet fraction ($f = N_{planets}/N_{stars}$) as a function
  of mass for our stellar sample (gray histogram). The red filled
  circles show the planet fraction predicted by
  Equation~\ref{eqn:model} for the masses and metallicities of the
  stars in each histogram bin. Note that we use a histogram only for
  visualization purposes; the data were fitted directly without
  binning. The open diamonds show the best-fitting 
  relationship between planet fraction and stellar mass for
  [Fe/H]~$ = 0$. The dashed line shows the stellar mass relationship
  predicted by \citet{kennedy08} for Solar
  metallicity. \label{fig:modelfit}}        
\end{figure*}

The agreement for the low-mass bin is not too surprising since we are
using the same sample of M dwarfs as used by Johnson et al. The
disagreement for the FGK dwarfs is in part due to the different
selection criteria for planet detections: we use a velocity amplitude
cutoff of $K > 20$~\ms, compared to the \msini~$ > 0.8$~\mjup\ used by
Johnson et al. Because of this, our sample of planet detections
includes a larger 
number of low-mass planets at short orbital periods, particularly 
for the Sun-like stellar sample. At higher stellar
masses, our measured planet fraction represents a significant
refinement over the result of Johnson et al., which stems primarily
from our larger sample size and higher Doppler precision with
Keck/HIRES compared to Lick/Hamilton. 

Our revised planet fraction for the high-mass stars appears much
smaller than the recent results presented by \citet{bowler10}, who
measured  $f = 26^{+9}_{-8}$\% for $1.5 \leq M_\star/M_\odot < 1.9$,
based on the Lick subgiants sample. However, Bowler et al. reported
the bulk occurrence rate, and did not attempt to correct or fit for
metallicity. Our analysis shows that metallicity plays an
important role in shaping the bulk occurrence rate among our
subgiants, which are metal-rich by $+0.14$~dex compared to
the less massive stars. This can be seen in the highest mass bin in
Figure~\ref{fig:modelfit}, in which the measured planet fraction is
consistent with the value measured by Bowler et al. Similarly, in
their analysis of the planet fraction for M dwarfs \citep{johnson10a}
noted the higher occurrence for metal-rich M dwarfs, but only reported
a bulk occurrence rate for the sample.

\subsection{Is there a planet-metallicity correlation among our evolved stars?}

In their analysis of the metallicity distribution of K giants with
planets, \citet{pasquini07} concluded there was no evidence of a PMC
among their evolved stars. \citet{takeda08} found a similar result
based on their sample of massive K giants, while \citet{hekker07}
did find evidence supporting a PMC among their stellar
sample. This somewhat contentious point has important implications for
the interpretation of the PMC seen among Sun-like stars. Pasquini et
al. argued that this lack of a metallicity  
correlation was evidence for the ``pollution'' scenario, in which only
the outer layers of stars with planets were metal-enriched by the
infall of gas-depleted planetesimals during the planet-formation epoch
\citep{gonzalez97, murray02}. In this 
scenario, as stars evolve off of the main sequence their convective
envelopes deepen and their polluted outer layers are diluted, erasing
any ``skin-deep'' metallicity enhancement \citep{laughlin00}. 

In our analysis, we implicitly assume that the PMC holds among the 
evolved stars in our sample. We can test this assumption by
restricting our analysis to $M_\star > 1.4$~\msun\ and comparing
the Bayesian evidence between two models: planet fraction as a function
of stellar mass alone, $f(M) = C M^\alpha$ (corresponding to a flat
metallicity distribution) versus planet fraction as a
function of stellar mass and metallicity, $f(M,F)$, given by
Equation~\ref{eqn:model}. We fitted both 
models to the subsample of massive subgiants, which
have deep convective envelopes according to the Padova stellar model
grids \citep{girardi02}. We find that the planet fraction among these 
evolved stars is best described as a function of mass and
metallicity, with an evidence ratio of order $10^{12}$. 

The extreme magnitude of the Bayes factor is driven primarily by the 7 
subgiants with $M_\star > 1.4$~\msun\ and [Fe/H]~$> +0.35$
(Figure~\ref{fig:massmet}). It is highly improbable that a flat
metallicity distribution would result in 5 out of 7 of these
metal-rich subgiants harboring a planet. The best fitting
parameters are $\alpha = 1.5 \pm 0.4$ and $\beta = 0.73 \pm 0.35$,
which are lower than, yet consistent with the values we measure for
the full stellar 
sample. However, the size and metallicity range of our sample of
subgiants only allows us to 
rule out a flat metallicity relationship with 98\% confidence. {\it At
present we can say that our data are consistent with a PMC among our
massive subgiants.}

We are not certain about the source of disagreement between our result
and those of Pasquini et al. and Takeda et al. One possibility is the
difference in our statistical methodologies. Both of those previous
studies compared the histograms of stars with and without
planets, and as a result did not quantify their confidence in  
a PMC or lack thereof. It will be informative to 
apply the techniques outlined in \S~\ref{fraction} to the stellar and
planet samples of the various K-giant surveys in order to make a
meaningful comparison with our results. 

\section{Summary and Discussion}
\label{summary}

We have used a large sample of planet-search target stars and planet
detections from the CPS to study the correlation between stellar
properties and the occurrence of giant planets ($K > 20$~\ms) with
$a \lesssim$2.5~AU. We have derived an
empirical relationship describing giant planet occurrence as a
function of stellar mass and metallicity, given by 

\begin{eqnarray}
  f(M_\star,{\rm [Fe/H]}) &=& \cbest \pm 0.01
  \times (M_\star/M_\odot)^{\abest \pm 0.3} \nonumber \\ 
  &\times& 10^{\bbest \pm 0.2 \rm [Fe/H]}. 
\end{eqnarray}

Our understanding of planet formation is presently dominated by two
theories: core accretion \citep[e.g.][]{pollack96} and disk
instability \citep{boss97}. The core accretion
model is a bottom-up process, by which 
protoplanetary cores are built up by the collisions of smaller
planetesimals. Once the core reaches a critical mass of roughly
10~$M_\oplus$, it rapidly accretes gas from the surrounding disk
material. The disk instability
mechanism is a top-down process whereby giant planets form from the
gravitational collapse of an unstable portion of the protoplanetary
disk. 

Both models depend on the existence of a massive gas disk, a portion
of which forms the bulk of the final mass of the Jupiter-like
planet. However, since the inner gas disks of protoplanetary disks
disperse on timescales of 3-5~Myr, the process of planet formation is
a race against time \citep[e.g.][]{pickett04}. In this race the disk
instability model holds 
a major advantage over the core accretion model because, under the
right conditions, planets can form from disk collapse in a mere
thousands of years, compared to of order Myr timescales required for
core accretion.  

The disk instability model predicts that there should
be no dependence on planet formation and physical stellar
properties. The simulations of \citet{boss06} showed that giant 
planets should readily form in even low-mass protoplanetary disks, and
that in general giant planets should form efficiently via disk
instability independent of stellar mass. The disk instability model
also predicts 
that planet formation should also be independent of disk metallicity
\citep{boss02}. Indeed, \citet{cai06} and \citet{meru10} show that the
efficiency of disk instability to form giant planets {\it decreases}
with increasing metallicity. 

In contrast to these predictions of the disk instability model, we
observe a strong dependence between planet occurrence and the physical
properties of the star. Assuming that the present-day mass and
metallicity of a star reflects the conditions in its protoplanetary
disk, then our results suggest that disk instability is not the primary
formation mechanism for the giant planets detected by Doppler
surveys. Indeed, it has long been recognized in that there are
theoretical complications in forming close-in planets via disk
instability \citet[][e.g.]{boley09}. However, the mechanism may be
responsible for planets in wide orbits \citep{dr09}, however see
\citet{kratter10} for complications with this scenario. 

An alternative explanation for the observed planet-metallicity
correlation is the so-called ``pollution'' model. In this scenario,
planet formation actually occurs around stars of all metallicities,
and the accretion of gas-depleted protoplanets onto the thin
convective layers of stars gives rise to an
enhanced stellar metallicity that is actually only "skin deep"
\citep{murray02}.
\citet{pasquini07} interpreted the flat metallicity distribution of K
giants with planets as evidence for such an effect, since the
deepening convective envelopes of evolved stars should dilute any
metallicity enhancement of its outer layers. However, we find that the
planet fraction among our massive subgiants is described well by a
model with a monotonic rise as a function of both mass and
metallicity (Section~\ref{comparison}). 

While much attention has been given to evolved, intermediate-mass
stars in the investigation of the pollution paradigm
\citep[][FV05]{laughlin97}, stars at the other end of
the stellar mass scale provide another proving ground. M dwarfs 
have deep convective envelopes over their entire lifetimes, and stars
with masses below 0.4~\msun\ are expected to be completely
convective, at least in the absence of strong magnetic activity
\citep[e.g.][]{mullan01}. 

The evidence for a PMC among the subgiants,
together with a strong PMC seen among the M dwarfs, are highly
suggestive that the present-day metallicities of 
stars are representative of the compositions of their disks during the
planet-formation era. One compelling explanation for both the observed
PMC and stellar-mass correlation
is that the surface density of solids is a key
factor in the planet formation process \citep{laughlin04, ida04,
  robinson06}. If so, both higher stellar (disk) 
metallicity and higher stellar (disk) mass can generate the requisite
surface density for planet formation. 

The relationship between planet formation efficiency and stellar
mass/metallicity has been previously studied in the context the  core
accretion paradigm \citep{ikoma00,kornet05,kornet06,
  kennedy08, thommes08, kretke09,johansen09,mordasini09,dr09}. \citet{kennedy08} 
modeled the evolution of the temperature profile at the disk midplanes of
stars of various masses, and studied the width of the radial region
of disks in which protoplanetary cores form most efficiently. They
found that the disks around A-type stars on their descent to the main
sequence  have very broad
formation regions, and their models predicted a positive correlation
between stellar mass and giant planet formation efficiency. Their
prediction for the planet fraction is shown as a dashed line in
Figure~\ref{fig:modelfit}, which we have approximated using the
polynomial relationship $f(M) = -0.03633 + 0.0138 M_\star -
0.0060 M_\star^2$. The agreement between theory and
observation is striking. 

The interplay between the mass and metallicity of protoplanetary
disks is also apparent in the core accretion simulations of
\citet{thommes08}. In their analysis, they simulated disks with a wide
variety of masses, viscosities and metallicities. Their models produce gas
giants most effectively in disks with a combination of high masses
($M_{\rm disk} \gtrsim 0.04$~\msun) and low viscosities. In their
simulations of 
the effects of metallicity, they found that gas giants can form in
Solar-composition disks only if the disk masses exceed $\approx
0.06$~\msun, or twice the minimum-mass Solar nebula. This mass
threshold decreases to $\approx 0.03$~\msun\ for disks with [Fe/H]~$ =
0.25$. Thus, Thommes et al. showed that there can be a trade-off
between the mass and metallicity of a protoplanetary disk in forming
giant planets. In the core accretion paradigm, M dwarfs can form giant 
planets, but only if they have high metallicities. Similarly, even
low-metallicity A stars can form massive planets, owing to their more
massive disks \citep[see also][]{dr09}. 

Theories of planet formation have progressed in large
leaps thanks largely to the rapidly growing sample of exoplanets
discovered around other stars. Successful, generalized theories of the
origins of planetary systems must account for the observed
correlations between planet occurrence and stellar
properties. The findings presented herein are the result of more than
15 years of high-precision Doppler monitoring of nearby
stars. Additional information will soon pour in from other surveys
using techniques 
such as microlensing \citep[e.g.][]{dong09,gould10}, astrometry
\citep{boss09}, transits \citep{kepler,corot1,irwin09} and direct 
imaging \citep{sphere,nici,gpi}. 

Our results have important implications for the target selection of
these future planet surveys. We find that A-type 
stars harbor planets at an elevated rate compared to less massive
stars. However, it is not obvious whether A type stars make
the most promising targets for other types of surveys. Where massive
stars perhaps hold the most promise is for direct imaging surveys. The
first two planetary systems imaged around normal stars were both
young A-type dwarfs \citep{kalas08,marois08}. Are A dwarfs the ideal
direct imaging targets? The answer to this question will rely
on the results presented herein, along with a careful consideration of
the mass, metallicity, luminosity and age distributions of nearby
stars; and the orbital and physical properties of planets as a
function of stellar mass. This issue will be addressed in a companion
paper (Crepp \& Johnson 2010). 

\acknowledgements

We gratefully acknowledge Debra Fischer, Geoff Marcy and Brendan
Bowler for their helpful suggestions, edits and comments. We  
thank Michael Fitzgerald, Jon Swift, Michael  
Cushing, Jason Wright and Brendan Bowler for their
illuminating discussions about Bayesian inference and other
statistical methods. Thanks to Scott Kenyon, Sally Dodson-Robinson and
Christoph Mordasini for their comments on earlier drafts of this
paper, and we acknowledge the careful edits and thoughtful suggestions
of the anonymous referee. JAJ acknowledges the NSF Astronomy and  
Astrophysics Postdoctoral Fellow with support from the NSF grant
AST-0702821 for supporting the research leading up to this work. KMA's
research was supported by the University of Hawaii 
Institute for Astronomy Research Experiences for Undergraduates (REU)
Program, which is funded by the National Science Foundation through
the grant AST-0757887. A.W.H. gratefully acknowledges support
from a Townes Post-doctoral Fellowship at the U.\,C.\ Berkeley Space
Sciences Laboratory. This publication makes use of 
the SIMBAD database operated at CDS, Strasbourge, France;
NASA's Astrophysics Data System Bibliographic Services; and the
Exoplanet Data Explorer at {\tt exoplanets.org}.

\bibliography{}

\end{document}